\documentclass[sigconf]{acmart}
\AtBeginDocument{%
  \providecommand\BibTeX{{%
    \normalfont B\kern-0.5em{\scshape i\kern-0.25em b}\kern-0.8em\TeX}}}

\usepackage{algorithmic}
\usepackage{graphicx}
\usepackage{textcomp}
\usepackage{xcolor}
\usepackage{array}
\acmConference[ESEC/FSE 2023]{The 31st ACM Joint European Software Engineering Conference and Symposium on the Foundations of Software Engineering}{03 - 09 December, 2023}{San Francisco, USA}
\begin{document}

\title{Revolutionizing API Documentation through Summarization}

\author{AmirHossein Naghshzan}
% \authornote{The academic advisor is Professor Sylvie Ratté.}
\email{amirhossein.naghshzan.1@ens.etsmtl.ca}
\affiliation{%
  \institution{École de Technologie Supérieure}
  \city{Montreal}
  \state{Quebec}
  \country{Canada}
}

\author{Sylvie Ratté}
\email{sylvie.ratte@etsmtl.ca}
\affiliation{%
  \institution{École de Technologie Supérieure}
  \city{Montreal}
  \state{Quebec}
  \country{Canada}
}


\begin{abstract}
This study tackles the challenges associated with interpreting Application Programming Interface (API) documentation, an integral aspect of software development. Official API documentation, while essential, can be lengthy and challenging to navigate, prompting developers to seek unofficial sources such as Stack Overflow. Leveraging the vast user-generated content on Stack Overflow, including code snippets and discussions, we employ BERTopic and extractive summarization to automatically generate concise and informative API summaries. These summaries encompass key insights like general usage, common developer issues, and potential solutions, sourced from the wealth of knowledge on Stack Overflow. Software developers evaluate these summaries for performance, coherence, and interoperability, providing valuable feedback on the practicality of our approach.
\end{abstract}





\maketitle

\section{Introduction}
The growing availability of textual data, especially in software development, offers both challenges and opportunities for efficient information extraction. This is particularly relevant in the context of complex Application Programming Interface (API) documentation, which is essential for programmers but can be lengthy, intricate, and occasionally incomplete. Official documentation, while valuable, can also be time-consuming to navigate \cite{Ponzanelli}.

Developers often turn to unofficial sources like Stack Overflow and GitHub for quicker access to information, highlighting the need for a more user-friendly approach to extracting insights from API documentation. To address this, our research introduces a novel method that utilizes BERTopic \cite{bertopic}, a topic modeling technique that combines BERT embeddings \cite{bert} and c-TF-IDF, for topic extraction and identifies common problems and solutions on Android APIs as a case study. Furthermore, we use extractive summarization to generate summaries for these topics and provide insightful documentation.

Our approach involves data gathering, preprocessing, employing BERTopic modeling for topic extraction, and utilizing extractive summarization. We assess the resulting summaries and topics for their performance, coherence, and interpretability, all with the objective of making API and method comprehension easier for developers.

To guide our research, we aim to answer the following research questions:

\vspace{0.2cm}

\textbf{RQ1}: {What are the prevalent topics discussed on Stack Overflow related to Android APIs?}

\vspace{0.2cm}

\textbf{RQ2}: {Can summarization methods effectively identify common issues within these topics?}

\vspace{0.2cm}

\textbf{RQ3}: {Is it feasible to derive solutions for these common issues by leveraging information from unofficial documentation sources?}

\vspace{0.2cm}

The goal is to assist developers in understanding relevant API methods for their daily tasks. This research could lead to valuable tools like IDE plugins or recommender systems for practical use by researchers and developers.

\section{Related Work}
The field of automatic summarization has gained significant attention, especially in code summarization. However, there's a gap in applying summarization techniques to APIs, especially with unofficial documentation sources. For instance, Alhaj {\em et al.}~\cite{Alhaj} used BERTopic to classify cognitive distortions in Arabic Twitter content, while Egger {\em et al.}~\cite{Egger} evaluated topic modeling techniques for social science research using Twitter data.

% The field of automatic summarization has seen significant research interest, particularly in the context of code summarization. However, there has been a noticeable gap in the application of summarization techniques to APIs, particularly when leveraging unofficial documentation sources. For instance, Alhaj {\em et al.}~\cite{Alhaj} tackled the challenge of classifying cognitive distortions in Arabic content on Twitter by employing BERTopic, a topic modeling approach that uncovered latent topics within tweets, resulting in improved text representations. In a similar vein, Egger {\em et al.}~\cite{Egger} conducted a comprehensive evaluation of topic modeling techniques using Twitter data for social science research, emphasizing the importance of choosing the right technique based on dataset characteristics.

Shifting the focus to code summarization, Sridhara {\em et al.}~\cite{Sridhara} demonstrated the effectiveness of Natural Language Processing (NLP) in accurately summarizing Java methods, preserving critical information in the process. Meanwhile, Abdalkareem {\em et al.}~\cite{Abdalkareem} devised an approach to extract code examples from Stack Overflow posts using HTML encoding and specialized tags filtering, a method found to be effective and adopted in our study. Hu {\em et al.}~\cite{Hu} introduced DeepCom, a system using NLP to generate code comments from large codebases, effectively handling challenges like extracting keywords from poorly named methods. Naghshzan {\em et al.}~\cite{naghshzan,naghshzan2} innovatively utilized Stack Overflow discussions to produce natural language summaries, providing developers with a valuable additional resource and highlighting the significance of unofficial documentation.

\section{Methodology}

\begin{table*}
  \centering
\caption{Top 3 topics of Android posts in Stack Ovberflow}
\renewcommand{\arraystretch}{1.4}% Spread rows out...
\begin{tabular}{c|c|l|l}
 \hline
 \textbf{Topic} & \textbf{Count} & \textbf{Name} & \textbf{Representation}\\
 \hline
   1 & 14663 & project\_error\_build\_gradle & project, proguard, studio, error, build, library, file, gradle, android, eclipse \\ \hline
   2 & 13471 & fragment\_viewpager\_view & fragment, recyclerview, item, view, listview, scroll, adapter, list, layout, row \\ \hline
   3 & 9947 & notification\_activity\_service\_gcm & notification, activity, service, gcm, app, analytics, push, back, intent, broadcast \\ \hline
   % 4 & 7752 & image\_camera\_bitmap\_opencv & image, camera, bitmap, opencv, screen, size, gallery, picasso, wallpaper, picture \\ \hline
   % 5 & 6474 & bar\_drawer\_navigation\_menu & bar, drawer, navigation, menu, progress, action, seekbar, toolbar, actionbar, layout \\ \hline
   % 6 & 6425 & keyboard\_search\_edittext\_text & keyboard, search, edittext, text, searchview, input, focus, textview, soft, autocompletetextview \\ \hline
   % 7 & 5950 & bluetooth\_device\_connection\_connect & bluetooth, device, connection, connect, ble, wifi, usb, android, connected, socket \\ \hline
   % 8 & 5372 & animation\_tab\_fragment\_viewpager & animation, tab, fragment, viewpager, view, transition, page, animate, activity, pager \\ \hline
   % 9 & 5320 & play\_google\_purchase\_app & location, gps, sensor, accelerometer, place, get, service, distance, latitude, google \\ \hline
   % 10 & 4249 & location\_gps\_sensor\_accelerometer & database, sqlite, table, db, query, room, column, cursor, data, insert \\
   %  \hline
\end{tabular}
  \label{topics}
\end{table*}
Our research methodology consists of three main steps: data collection, topic modeling, and summarization. In the following section, we delve into a comprehensive examination of these key processes.

\subsection{Data Collection and Pre-processing}
In the data collection phase, we retrieved all Stack Overflow questions tagged with {\em Android} from January 2009 to April 2022, totaling 3,698,168 posts. To focus on natural language for topic modeling and summarization, we removed code blocks highlighted by the HTML tag (\textless code\textgreater). Data preprocessing steps included eliminating stop words, punctuation, tokenization, lemmatization, stemming, and removing special characters and numerals.

\subsection{Topic Modeling}
To identify topics in Stack Overflow discussions, we employed BERTopic \cite{bertopic}, a powerful topic modeling technique based on BERT, a pre-trained transformer model, for enhanced semantic understanding. Traditional techniques like LDA \cite{Jelodar} struggle with semantic context \cite{Egger}. BERTopic overcomes this by integrating BERT's contextual information. We used a pre-trained model available on Hugging Face\footnote{https://huggingface.co} and Google Colab Pro's T4 GPU for efficient computations. 

\subsection{Summarization}
Previously, Naghshzan \emph{et al.}~\cite{naghshzan} demonstrated that extractive summarization can be employed to extract information from Stack Overflow. Therefore, we utilize the extractive version of BERT \cite{newbert}. 

% In extractive summarization, BERT assigns importance scores to sentences or phrases within the input document, and the highest-scoring sentences or phrases are selected to form the summary

Our summarization approach involves two key steps:\\
\textbf{Summarizing Topics Based on Questions:} In the first step, we generate summaries for each topic using questions. We believe that summarizing topics through questions can help identify the most prevalent issues within each topic. By focusing on the top topics from Stack Overflow, we aim to uncover the primary challenges in Android development discussed on the platform.\\
\textbf{Summarizing Topics Based on Answers:} The second step involves creating summaries of topics based on the corresponding answers. These summaries provide potential solutions to the challenges identified in the previous step, drawing from the collective wisdom of the Stack Overflow community.

To implement this, we tokenize the input document with the BERT tokenizer and feed the tokens into the BERT model, which encodes contextual information and captures semantic meaning. The summarization algorithm, based on BERT, calculates importance scores for each sentence or phrase, selecting the highest-scoring ones for the summary.

In the first step, we use BERT to generate summaries of questions related to identified topics. The BERTopic algorithm provides us with relevant posts for each topic, and we use this data to create summaries.

Moving to the second step, we seek solutions to the identified problems. We locate related questions in the database and note their IDs. Using these IDs, we find answers associated with the questions and apply the same summarization process to generate answer summaries. To ensure quality, we select answers marked as accepted or with a score above the average, typically a score of 2 or higher.

\section{Results}

We identified 81 topics for Android discussion in Stack Overflow. Table \ref{topics} showcases the three most prevalent topics related to Android posts on Stack Overflow. The \textit{Count} column quantifies the number of posts associated with each specific topic. The \textit{Name} column contains the names of the topics, as generated by BERTopic. The \textit{Representation} column lists the words that best represent each respective topic. This table offers a concise summary of the data, providing insights into the topics that most frequently dominate Android-related discussions on the platform. A comprehensive list of our findings can be accessed in the online Appendix\footnote{https://github.com/amirarcane/BERTopic}.

\begin{table*}
  \centering
\caption{Generated summaries for questions and answers of topic project\_error\_build\_gradle}
\renewcommand{\arraystretch}{1.4}% Spread rows out...
\begin{tabular}{m{8.5cm} | m{8.5cm}}
 \hline
 \textbf{Questions} & \textbf{Answers}\\
 \hline
   Jenkins tries to launch tools instead of emulator I'm trying to set up jenkins ui tests and it fails on running emulator command & I'm answering my own question its an issue with android emulator plugin not working with new command line tools only sdk package  \\ \hline
   I am trying to add kotlin sources of an aar in android studio it doesnt work when I select choose sources and choose the corresponding source jar & add classpath org.jetbrains.kotlin gradle plugin clean and build your project it worked for me  \\ \hline
   I have an sdk works with google api and implementing a lot of dependencies then I implement the lib into my new app which is also implements dependencies everything goes fine until I try to run the app & the problem is that you have a duplicated dependency with different version if you can find the implementation that needs com.google.android.gms.location you can put under it exclude com.google.android.gms or simply in build.gradle \\ \hline
\end{tabular}
  \label{summaries}
\end{table*}

Table \ref{summaries} presents sample summaries for the top two topics, highlighting three common problems for each topic. Table \ref{summaries} presents generated answer summaries for the previously identified problems, offering potential solutions. For example, for the first topic, one common problem is \textit{"jenkins tries to launch tools emulator instead of emulator"} and the generated summary suggests a possible solution: \textit{"its an issue with android emulator plugin not working with new command line tools only sdk package"}.

\section{Evaluation}
We conducted a user study with 15 software developers, including both experienced and intermediate-level programmers, to evaluate our API documentation summarization approach. Participants rated the summaries for coherence, informativeness, relevance, and user satisfaction. The results show moderately positive outcomes with average scores of 3.8 for coherence, 3.7 for informativeness, 3.9 for relevance, and 3.6 for user satisfaction. These results indicate the potential of our approach to assist developers in understanding complex API documentation.

\section{Discussion}
In our research, we tackled the challenges posed by complex API documentation by introducing an innovative approach for generating API summaries from informal sources like Stack Overflow. Our methodology involved BERTopic for topic modeling and BERT for text summarization.

To address our first research question (RQ1), we identified recurring topics discussed on Stack Overflow concerning Android APIs. We applied BERTopic to a dataset of 3,698,168 unique Android posts, extracting 80 prominent topics that shed light on frequently discussed issues in Android development.

Moving on to our second research question (RQ2), we employed text summarization techniques to identify common issues within these topics. By summarizing questions related to each topic, we pinpointed prevalent problems gleaned from the Stack Overflow community's collective knowledge.

Lastly, addressing our third research question(RQ3), we harnessed information from unofficial documentation to generate solutions to these common issues. Summarizing answers to related questions, we provided potential solutions sourced from the Stack Overflow community's insights, aiming to address identified challenges.

To wrap up, our research demonstrated the feasibility of automatically generating API summaries from informal sources, streamlining the understanding of APIs and methods for developers. While this study offers valuable insights and serves as a foundational exploration, further research and validation are essential to refine and expand upon these findings. There is ample room for improvement and future investigations in this area.

\balance

\section{Conclusion and Future Work}
In this work, we introduce an automated method for generating API summaries from informal sources like Stack Overflow. Utilizing BERTopic and NLP techniques, we extract recurring topics, condensing common problems and their solutions. This approach alleviates developers' challenges in navigating complex API documentation, streamlining the process. By automatically extracting and summarizing vital insights, our method saves developers valuable time, categorizing common development issues and their resolutions.

As we continue to refine and expand our approach, there are several avenues for future work that can enhance and expand upon our findings. The following are potential areas for future research: \\
\textbf{Extension to Diverse Domains:} Broaden the applicability and impact of our work by expanding our approach beyond Android APIs to encompass various domains and programming languages. \\
\textbf{Exploring Multiple Information Sources:} Provide a
more comprehensive and diverse set of data by investigating alternative platforms like GitHub, bug reports, blogs, and forums. \\
\textbf{Investigating Summarization Techniques:} Exploring various summarization approaches, including extractive, abstractive, and hybrid methods to determine their suitability for API summarization tasks. \\
\textbf{Integration with Developer Tools:} Seamlessly integrating API summaries into developer tools and IDEs to enhance productivity. 

In conclusion, our method has the potential to assist developers in comprehending and using APIs more effectively. By pursuing these future avenues, we aim to advance developer productivity and software engineering practices.

% \appendix




\end{document}